\documentclass[12pt]{article}
\begin{document}
\title{STATIC CYLINDRICAL SYMMETRY AND CONFORMAL FLATNESS}

\author{L. Herrera$^{1}$\thanks{e-mail: laherrera@telcel.net.ve},
G. Le Denmat$^{2}$\thanks{e-mail: gele@ccr.jussieu.fr}, G. Marcilhacy$^{2}$
and N. O. Santos$^{2,3,4}$\thanks{e-mail: santos@ccr.jussieu.fr and
nos@cbpf.br} \\
{\small $^{1}$Escuela de F\'{\i}sica, Facultad de Ciencias,}\\
{\small Universidad Central de Venezuela, Caracas, Venezuela.}\\
{\small $^{2}$Universit\'{e} Pierre et Marie Curie - CNRS/FRE 2460,}\\
{\small LERMA/ERGA, Tour 22-12, 4\'eme \'etage, Bo\^{\i}te 142, 4 place
Jussieu,}\\
{\small 75252 Paris Cedex 05, France.}\\
{\small $^{3}$Laborat\'{o}rio Nacional de Computa\c{c}\~{a}o Cient\'{\i}fica,}\\
{\small 25651-070 Petr\'opolis RJ, Brazil.}\\
{\small $^{4}$Centro Brasileiro de Pesquisas F\'{\i}sicas,}\\
{\small 22290-180 Rio de Janeiro RJ, Brazil.}}
\maketitle

\begin{abstract}
We present the whole set of equations with regularity and matching
conditions required for the description of physically meaningful static
cylindrically symmmetric distributions of
matter, smoothly matched to Levi-Civita vacuum spacetime. It is shown
that the conformally flat solution with equal principal stresses represents
an incompressible fluid.
It is also proved that any conformally flat cylindrically symmetric static
source cannot be matched through Darmois conditions to the Levi-Civita
spacetime. Further evidence is given
that when the Newtonian mass per unit length reaches 1/2 the spacetime has
plane symmetry.
\end{abstract}

\newpage
\section{Introduction}
Cylindrical systems in Einstein's theory puzzles relativists since
Levi-Civita found its vacuum solution in 1919 \cite{Levi}. The precise
meaning of its two independent parameters
 is still hard to grasp and, in particular, the one that describes the
Newtonian energy per unit length $\sigma$ looks the most elusive. The fact
that there are two parameters while in
its counterpart, Newtonian theory, has only one parameter looks a
sufficient justification for deserving more research. But the importance of
its research goes further if one notices
the close link between Levi-Civita, $\gamma$ and Schwarzschild spacetimes \cite{Herrera4}
and its peculiar properties.
Besides, there has been renewed interest in cylindrically symmetric sources
in relation with different, classical and quantum, aspects of gravitation
(see \cite{1} and references
therein).  Such sources may serve as test--bed for numerical relativity,
quantum gravity and for probing cosmic censorship and hoop conjecture,
among other important issues, and
represent a natural tool to seek the physics that lies behind the two
independent parameters in Levi-Civita metric.

The purpose of this work is twofold. On the one hand we would like to
present systematically the field equations as well as all regularity and
junction conditions required to ensure
the correct behaviour of a source of a cylindrically symmetric spacetime
(Levi-Civita). On the other hand we want to bring out the relationship
between the Weyl tensor and different
aspects of the source. This last question is in turn motivated by the very
conspicuous link  existing in the spherically symmetric case between the
Weyl tensor, the inhomogeneity of
the energy density and the anisotropy of pressure \cite{est}.

The paper is organized as follows: in section 2 we present the general form
of the energy momentum tensor, the line element, the Einstein equations,
the active gravitational mass and
the Weyl tensor. The exterior space-time as well as junction and
regularity conditions are discussed in section 3. In section 4 the
consequences derived from the condition of conformal
flatness are obtained. The non existence of conformally flat models
satisfying Darmois conditions is given in section 5. Finally, some
conclusions are presented in the last section.
\section{Interior spacetime}
We consider a static cylindrically symmetric anisotropic non-dissipative
fluid bounded by a cylindrical surface $\Sigma$ and with energy momentum
tensor given by
\begin{equation}
T_{\alpha\beta}=(\mu + P_r)V_{\alpha}V_{\beta}+P_rg_{\alpha\beta}+
(P_{\phi}-P_r)K_{\alpha}K_{\beta}+(P_z-P_r)S_{\alpha}S_{\beta}, \label{1}
\end{equation}
where, $\mu$ is the energy density, $P_r$, $P_z$ and $P_{\phi}$ are the
principal stresses and $V_{\alpha}$, $K_{\alpha}$ and $S_{\alpha}$ are
vectors satisfying
\begin{equation}
V^{\alpha}V_{\alpha}=-1, \;\; K^{\alpha}K_{\alpha}=S^{\alpha}S_{\alpha}=1, \;\;
V^{\alpha}K_{\alpha}=V^{\alpha}S_{\alpha}=K^{\alpha}S_{\alpha}=0. \label{2}
\end{equation}
We assume for the interior metric to $\Sigma$ the general static
cylindrically symmetric which can be written
\begin{equation}
ds^2=-A^2dt^2+B^2(dr^2+dz^2)+C^2d\phi^2, \label{3}
\end{equation}
where $A$, $B$ and $C$ are all functions of $r$. To represent cylindrical
symmetry, we impose the following ranges on the coordinates
\begin{equation}
-\infty\leq t\leq\infty, \;\; 0\leq r, \;\; -\infty<z<\infty, \;\;
0\leq\phi\leq 2\pi. \label{3a}
\end{equation}
We number the coordinates $x^0=t$, $x^1=r$, $x^2=z$ and $x^3=\phi$ and we
choose the fluid being at rest in this coordinate system, hence from
(\ref{2}) and (\ref{3}) we have
\begin{equation}
V_{\alpha}=-A\delta_{\alpha}^0, \;\; S_{\alpha}=B\delta_{\alpha}^2, \;\;
K_{\alpha}=C\delta_{\alpha}^3. \label{4}
\end{equation}

For the Einstein field equations, $G_{\alpha\beta}=\kappa T_{\alpha\beta}$
with (\ref{1}), (\ref{3}) and (\ref{4}) we have the non null components
\begin{eqnarray}
G_{00}=-\left(\frac{A}{B}\right)^2\left[
\left(\frac{B^{\prime}}{B}\right)^{\prime}+\frac{C^{\prime\prime}}{C}\right]
=\kappa\mu A^2, \label{5} \\
G_{11}=\frac{A^{\prime}}{A}\frac{C^{\prime}}{C}+\left(\frac{A^{\prime}}{A}+\frac{C^{\prime}}{C}
\right)\frac{B^{\prime}}{B}=\kappa P_rB^2, \label{6} \\
G_{22}=\frac{A^{\prime\prime}}{A}+\frac{C^{\prime\prime}}{C}+\frac{A^{\prime
}}{A}
\frac{C^{\prime}}{C}-\left(\frac{A^{\prime}}{A}+\frac{C^{\prime}}{C}\right)\frac{B^{\prime}}{B}
=\kappa P_zB^2, \label{7} \\
G_{33}=\left(\frac{C}{B}\right)^2\left[\frac{A^{\prime\prime}}{A}+
\left(\frac{B^{\prime}}{B}\right)^{\prime}\right]=\kappa P_{\phi} C^2, \label{8}
\end{eqnarray}
where the primes stand for differentiation with respect to $r$. Since we
have four equations for seven unknown functions, three additional
constraints (e.g. equations of state) should
be given in order to uniquely determine a solution.

There are two compact expressions that can be obtained from (\ref{6}-\ref{8}),
\begin{eqnarray}
\kappa(P_r+P_z)B^2=\frac{(AC)^{\prime\prime}}{AC}, \label{8d} \\
\kappa(P_z-P_{\phi})B^2=\frac{h^{\prime\prime}}{h}+\left(\frac{A^{\prime}}{A}+
\frac{B^{\prime}}{B}\right)\frac{h^{\prime}}{h}, \label{8e}
\end{eqnarray}
where
\begin{equation}
h=\frac{C}{B}. \label{8f}
\end{equation}

The conservation equation, $T_{r;\beta}^{\beta}=0$, with (\ref{1}) and
(\ref{3}) becomes
\begin{equation}
(\mu+P_r)\frac{A^{\prime}}{A}+P_r^{\prime}+(P_r-P_z)\frac{B^{\prime}}{B}+(P_
r-P_{\phi})
\frac{C^{\prime}}{C}=0, \label{8aa}
\end{equation}
which can substitute any of the independent field equations (\ref{5}-\ref{8}).

The Whittaker formula \cite{Whittaker} for the active gravitational mass
per unit length $m$ of a static
distribution of perfect fluid with energy density $\mu$ and principal
stresses $P_r$, $P_z$ and
$P_{\phi}$ inside a cylinder of surface $\Sigma$ is
\begin{equation}
m=2\pi \int_0^{r_{\Sigma}}(\mu +P_r +P_z +P_{\phi})\sqrt{-g}dr, \label{8a}
\end{equation}
where $g$ is the determinant of the metric. Now substituting (\ref{3}) and
(\ref{5}-\ref{8}) into (\ref{8a}) we obtain
\begin{equation}
m=\frac{4\pi}{\kappa}\int_0^{r_{\Sigma}}\left(\frac{A^{\prime\prime}}{A^{\prime}}+
\frac{C^{\prime}}{C}\right)A^{\prime}C\;dr, \label{8b}
\end{equation}
which can be recast into the simpler form
\begin{equation}
m=\frac{4\pi}{\kappa}\int_0^{r_\Sigma}(A^{\prime}C)^{\prime}dr. \label{8c}
\end{equation}

The spacetime (\ref{3}) has the following non-null components of the Weyl
tensor $C_{\alpha\beta\gamma\delta}$
\begin{eqnarray}
C_{1212}=-\left(\frac{B^2}{AC}\right)^2C_{0303}=\frac{B^2}{6}\left[\frac{A^{\prime\prime}}{A}-
2\left(\frac{B^{\prime}}{B}\right)^{\prime}+\frac{C^{\prime\prime}}{C}-
2\frac{A^{\prime}}{A}\frac{C^{\prime}}{C}\right], \label{9} \\
C_{1313}=-\left(\frac{C}{A}\right)^2C_{0202} \nonumber \\
=\frac{C^2}{6}\left[\frac{A^{\prime\prime}}{A}+
\left(\frac{B^{\prime}}{B}\right)^{\prime}-2\frac{C^{\prime\prime}}{C}-
3\left(\frac{A^{\prime}}{A}-\frac{C^{\prime}}{C}\right)\frac{B^{\prime}}{B}+
\frac{A^{\prime}}{A}\frac{C^{\prime}}{C}\right], \label{10} \\
C_{2323}=-\left(\frac{C}{A}\right)^2C_{0101} \nonumber \\
=\frac{C^2}{6}\left[-2\frac{A^{\prime\prime}}{A}+
\left(\frac{B^{\prime}}{B}\right)^{\prime}+\frac{C^{\prime\prime}}{C}+
3\left(\frac{A^{\prime}}{A}-\frac{C^{\prime}}{C}\right)\frac{B^{\prime}}{B}+
\frac{A^{\prime}}{A}\frac{C^{\prime}}{C}\right]. \label{11}
\end{eqnarray}
We obtain from (\ref{9}-\ref{11})
\begin{equation}
\left(\frac{C}{B}\right)^2C_{1212}+C_{1313}+C_{2323}=0, \label{14}
\end{equation}
hence we have only two independent components of the Weyl tensor for (\ref{3}).

\section{Exterior spacetime and junction conditions}
For the exterior spacetime of the cylindrical surface $\Sigma$, since the
system is static, we take the Levi-Civita metric \cite{Levi},
\begin{equation}
ds^2=-a^2\rho^{4\sigma}dt^2+b^2\rho^{4\sigma(2\sigma-1)}(d\rho^2+dz^2)
+c^2\rho^{2(1-2\sigma)}d\phi^2, \label{16}
\end{equation}
where $a$, $b$, $c$ and $\sigma$ are real constants. The coordinates $t$,
$z$ and $\phi$ in (\ref{16}) can be taken the same as in (\ref{3}) and with
the same ranges (\ref{3a}).
The radial coordinates in (\ref{3}) and (\ref{16}), $r$ and $\rho$, are not
necessarily continuous on $\Sigma$ as we see below by applying the junction
conditions. The constants $a$
and $b$ can be removed by scale transformations, while $c$ cannot be
transformed away if we want to preserve the range of $\phi$ in (\ref{16})
\cite{16}. The constant $\sigma$
represents the Newtonian mass per unit length. (For a discussion of the
number of constants in cylindrical spacetimes see
\cite{Silva,MacCallum,Bonnor}.)

In accordance with the Darmois junction conditions \cite{Darmois}, we
suppose that the first fundamental form which $\Sigma$ inherits from the
interior metric (\ref{3}) must be the same as the one it inherits from the
exterior metric (\ref{16}); and similarly, the inherited second fundamental
form must be the same. The conditions are necessary and sufficient for a
smooth matching without a surface layer.

The equation of $\Sigma$, for the interior and exterior spacetimes, can be
written respectively as
\begin{equation}
f(r)=r-r_{\Sigma}=0, \;\; g(\rho)=\rho-\rho_{\Sigma}=0, \label{20}
\end{equation}
where $r_{\Sigma}$ and $\rho_{\Sigma}$ are constants. From (\ref{20}) we
can calculate the continuity of the first and second fundamental forms, and
we obtain,
\begin{eqnarray}
A_{\Sigma}=a\rho_{\Sigma}^{2\sigma},
\;\;B_{\Sigma}=b\rho_{\Sigma}^{2\sigma(2\sigma-1)},
\;\; C_{\Sigma}=c\rho_{\Sigma}^{1-2\sigma} \label{21} \\ 
\left(\frac{A^{\prime}}{A}\right)_{\Sigma}=\frac{2\sigma}{\rho_{\Sigma}}, \;\;
\left(\frac{B^{\prime}}{B}\right)_{\Sigma}=\frac{2\sigma(2\sigma-1)}{\rho_{\Sigma}}, \;\;
\left(\frac{C^{\prime}}{C}\right)_{\Sigma}=\frac{1-2\sigma}{\rho_{\Sigma}}.\label{22}
\end{eqnarray}

Considering (\ref{6}) on the surface $\Sigma$ and substituting into the
junction conditions (\ref{20}) we obtain
\begin{equation}
P_{r\Sigma}=0, \label{23}
\end{equation}
as expected.

The Whittaker mass per unit length (\ref{8c}) after integration and using
the junction conditions (\ref{21}) and (\ref{22}) becomes
\begin{equation}
m=\frac{4\pi}{\kappa}\left[2ac\sigma-(A^{\prime}C)_0\right], \label{23a}
\end{equation}
where the index 0 means the quantity evaluated at the axis of the mass
distribution.

Next, regularity conditions on the the axis of symmetry imply \cite{Philbin}

\begin{equation}
A^{\prime}(0)=B^{\prime}(0)=C^{\prime \prime}(0)=C(0)=0, \;\;\;
B(0)=C^{\prime}(0)=1,
\label{regul}
\end{equation}
 hence, considering the
gravitational coupling constant
$G=1$ then
$\kappa=8\pi$ and  (\ref{23a}) reduces to
\begin{equation}
m=ac\sigma. \label{23b}
\end{equation}
\section{Conformally flat interior}
The conformally flat condition imposes the vanishing of all Weyl tensor
components, hence from (\ref{9}-\ref{14}) we have
\begin{eqnarray}
S^{\prime}+S^2-\frac{2h^{\prime}}{h}S+\frac{h^{\prime\prime}}{h}=0,
\label{32} \\
S^{\prime}+S^2+\frac{h^{\prime}}{h}S-\frac{2h^{\prime\prime}}{h}=0,
\label{33}
\end{eqnarray}
where
\begin{equation}
S=\frac{A^{\prime}}{A}-\frac{B^{\prime}}{B}. \label{13}
\end{equation}
Then it follows
\begin{eqnarray}
h^{\prime}S-h^{\prime\prime}=0, \label{35n} \\
S^{\prime}+S^2-\frac{h^{\prime\prime}}{h}=0, \label{36n}
\end{eqnarray}
which produces
\begin{equation}
h^{\prime\prime\prime}-\frac{h^{\prime\prime}h^{\prime}}{h}=0. \label{37n}
\end{equation}
Let us now examine the two possible cases, $h^{\prime}\neq 0$ and $h^{\prime}= 0$

\subsection{Case $h^{\prime} \neq 0$}
We obtain from (\ref{37n}) after integration
\begin{equation}
h=a_1\exp(a_2r)+a_3\exp(-a_2r), \label{44n}
\end{equation}
where $a_1$, $a_2$ and $a_3$ are integration constants with the condition that
\begin{equation}
h^2\geq 4a_1a_3. \label{45}
\end{equation}
However, regularity conditions on the axis (\ref{regul}) require
\begin{equation}
a_1=-a_3, \label{39n}
\end{equation}
and (\ref{44n}) reduces to
\begin{equation}
h=a_1\sinh(a_2r), \label{40n}
\end{equation}
where $a_1$ was redefined.

Substituting (\ref{40n}) into (\ref{35n}) and integrating we have
\begin{equation}
A=a_3\cosh(a_2r)B, \label{46n}
\end{equation}
where $a_3$ is another integration constant.

Thus, conformal flatness reduce the total number
of unknown functions by two, through (\ref{40n}) and (\ref{46n}). However,
since the total number of
variables is seven, we
still need one condition in order to determine a solution uniquely.
So, let us consider the   three different cases of
isotropy.
\newpage 

i) $P_z=P_{\phi}$

Then we obtain from (\ref{7}), (\ref{8}) and (\ref{8f})
\begin{equation}
\frac{h^{\prime
\prime}}{h}+\frac{h^{\prime}}{h}\left(\frac{A^{\prime}}{A}+\frac{B^{\prime}}
{B}\right)=0,
\label{n2}
\end{equation}
which together with (\ref{35n}) yields that $A^{\prime}=0$, this in turn
implies, because of (\ref{46n}) and assuming without lost of generality $A=1$,
\begin{equation}
B=\frac{1}{\cosh(a_2r)}, \label{n3}
\end{equation}
where we chose $a_3=1$ to satisfy (\ref{regul}).
Feeding back (\ref{40n}), (\ref{46n}) and (\ref{n3}) into (\ref{5}--\ref{8})
we obtain
\begin{equation}
P_r=P_z=P_{\phi}=-\frac{\mu}{3}=-\frac{a_2^2}{\kappa}.
\label{n4}
\end{equation}
Thus the solution represents an incompressible cylinder with isotropic
(negative) stresses.

ii) $P_r=P_z$

From (\ref{6}) and  (\ref{7}),  we have
\begin{equation}
\frac{A^{\prime
\prime}}{A}+\frac{B^{\prime\prime}}{B}-2\left(\frac{A^{\prime}}{A}+
\frac{C^{\prime}}{C}\right)\frac{B^{\prime}}{B}=0,
\label{n5}
\end{equation}
and with (\ref{40n}) and (\ref{46n}) we obtain the equation for $B$,
\begin{equation}
\frac{B^{\prime\prime}}{B}-2\left(\frac{B^{\prime}}{B}\right)^2+a_2^2=0,
\label{n6}
\end{equation}
and by choosing the integration constants to satisfy (\ref{regul}) its
solution is
\begin{equation}
B=\frac{1}{\cosh(a_2r)}. \label{n6}
\end{equation}
From (\ref{46n}) and (\ref{n6}) and assuming $A=1$ this case yields the
same solution as the preceding one.

iii) $P_r=P_{\phi}$.

From (\ref{6}) and (\ref{8}), it follows
\begin{equation}
\frac{A^{\prime \prime}}{A}+\frac{B^{\prime
\prime}}{B}-2\left(\frac{B^{\prime}}{B}\right)^2-2\frac{A^{\prime}}{A}\frac{
B^{\prime}}{B}-\frac{h^{\prime}}{h}
\left(\frac{A^{\prime}}{A}+\frac{B^{\prime}}{B}\right)=0,
\label{n7}
\end{equation}
and substituting into it (\ref{40n}) and (\ref{46n}) leads to the equation
for $B$,
\begin{equation}
\frac{B^{\prime\prime}}{B}-2\left(\frac{B^{\prime}}{B}\right)^2-a_2\coth(a_2r)
\frac{B^{\prime}}{B}=0, \label{n8}
\end{equation}
which has the solution satisfying the regularity conditions (\ref{regul})
\begin{equation}
B=\frac{1}{a_4[\cosh(a_2r)-1]+1}, \label{n8}
\end{equation}
where $a_4$ is an integration constant.
Then substituting (\ref{n8}) into (\ref{46n}) we get
\begin{equation}
A=\frac{a_3\cosh(a_2r)}{a_4[\cosh(a_2r)-1]+1}
\label{n10}
\end{equation}

Using field equations (\ref{5}-\ref{8}) together with  (\ref{40n}),
(\ref{n8}) and
(\ref{n10}), we can obtain the expressions for the physical variables, which
are
\begin{eqnarray}
\kappa\mu=2a_2^2a_4\left[(1-a_4)\cosh(a_2r)+a_4-3\right]+3a_2^2, \label{n11} \\
\kappa P_r=\kappa
P_{\phi}=2a_2^2a_4\left[(a_4-1)\tanh(a_2r)\sinh(a_2r)+1\right]-a_2^2,
\label{n12} \\
\kappa P_z=2a_2^2a_4\left[\frac{1-a_4}{\cosh^2(a_2r)}+a_4+1\right]-a_2^2.
\label{n13}
\end{eqnarray}
Observe that in this case the matter distribution is not completely
isotropic in the stresses and the energy density is not homogeneous.

\subsection{Case $h^{\prime}=0$}
Then we have from (\ref{36n})
\begin{equation}
S=\frac{1}{b_1+r} \label{38n}
\end{equation}
where $b_1$ is an integration constant. Using (\ref{38n}) in (\ref{13}), we
obtain after
integration
\begin{equation}
A=Bb_2(b_1+r), \label{40nn}
\end{equation}
where $b_2$ is another integration constant.
However regularity conditions (\ref{regul}) imply from (\ref{40nn}) that
$A=0$, which is obviously unacceptable.
\vspace{1cm}

So far we have only assumed the spacetime to be conformally flat at the
interior, and regularity conditions to be satisfied. However as it can be
easily  checked, neither of the
models above satisfy the Darmois conditions (\ref{21}-\ref{22}). As a
matter of fact, and as it will be shown in the next section, there is no
conformally flat interior solutions
satisfying Darmois (and regularity) conditions.

\section{Non existence of conformally flat solution satisfying Darmois
conditions}
As we have seen if the cylinder has a matter content that is conformally
flat and satisfies
regularity  conditions on the axis then,
\begin{equation}
h=a_1\sinh(a_2r), \label{40nn}
\end{equation}
if $h^{\prime}\neq 0$.

Now considering the junction conditions (\ref{21}) and (\ref{22}) we obtain
from (\ref{40nn})
\begin{eqnarray}
a_1=\frac{c\rho_{\Sigma}^{1-4\sigma^2}}{b\sinh(a_2r_{\Sigma})}, \label{41} \\
a_2r_{\Sigma}\coth(a_2r_{\Sigma})=1-4\sigma^2. \label{42n}
\end{eqnarray}
Since always $a_2r_{\Sigma}\coth(a_2r_{\Sigma})>1$ then the condition
(\ref{42n}) can never be satisfied.

But if $h^{\prime}=0$, as we have seen before, regularityv conditions are not satisfied.

Hence we can state that {\it any static cylindrical source matched smoothly to
the Levi-Civita spacetime does not admit conformally flat solution}.

\section{Conclusions}
We have deployed the equations describing the static cylinder, as well as
the regularity and matching conditions. Then the consequences derived from
the assumption of conformal flatness
were obtained. It was shown that there exist no interior conformally flat
solution which satisfies regularity conditions and matches smoothly to
Levi-Civita spacetime on the
boundary surface. Of course if we relax Darmois conditions and allow for
the existence of shells at the boundary surface, the latter conclusion does
not hold.

It was also shown that the conformally flat, isotropic (in the stresses)
cylinder is necessarily incompressible ($\mu=$ constant). Inversely, since
the solution for the incompressible
isotropic cylinder is unique (there are four equations for four variables)
then it is clear that such solution is also conformally flat.

So, if we look for an incompressible cylinder matching smoothly to
Levi-Civita (hence not conformally flat), we have to relax the condition
of isotropy in the stresses . Thus for
example one could  assume $\mu=$ constant,
$P_z=P_{\phi}\neq P_r$, then we can integrate (\ref{8e}) to obtain
\begin{equation}
ABh^{\prime}=c_1,
\label{44n}
\end{equation}
where $c_1$ is an integration constant. By considering junction conditions
we get
\begin{equation}
c_1=ac(1-4\sigma^2).
\label{45n}
\end{equation}
From (\ref{23b}) and (\ref{45n}) it follows that as $\sigma \rightarrow
1/2$, $m \rightarrow \infty$. This result gives further evidence that the
spacetime at this limit for $\sigma$
has plane symmetry \cite{Bonnor,Philbin,Herrera,Herrera1}. Of course to
fully specify a solution another condition has to be given.

Finally it is worth noting the diferences and the similarities between this
case and the the spherically symmetric situation. For spherical symmetry
there is only one independent component of the Weyl tensor, while for
cylindrical symmetry there are two
independent components. For spherical symmetry the conditions of
incompressibility
and isotropic pressure lead also to a unique solution, the interior
Schwarzschild solution, which is conformally flat \cite{Raychau}, however
unlike our present
case, that solution can be matched
smoothly on the boudary surface to the exterior solution. If the condition
of isotropic pressure is relaxed in the spherically symmetric case,
conformally flat solutions matching
smoothly to Schwarzschild spacetime exist, but are not incompressible
\cite{H}. The same happens in the cylindrically symmetric case with
$P_r=P_{\phi} \neq P_z$, however in this case
the solution does not satisfy Darmois conditions.

\end{document}